\documentclass[tradiabstract]{aa}
\usepackage{graphicx}
\usepackage{txfonts}
\usepackage{natbib}
\bibpunct[; ]{(}{)}{;}{a}{}{,}

\begin{document}
\title{High Cadence Near Infrared Timing Observations of\\Extrasolar
  Planets: I. GJ\,436b and XO-1b\thanks{Based on observations collected at the
    European Southern Observatory, Chile. DDT project 279.C-5020, and
    079.C-0557.}}

\titlerunning{High Cadence Observations of GJ\,436b \& XO-1b.}

\author{C. C\'{a}ceres\inst{1,2}, V.D. Ivanov\inst{2},
  D. Minniti\inst{1,3}, D. Naef\inst{2}, C. Melo\inst{2},
  E. Mason\inst{2}, F. Selman\inst{2}, G. Pietrzynski\inst{4}}
\authorrunning{C\'{a}ceres et al.}

\offprints{C. C\'{a}ceres}

\institute{
  P. Universidad Cat\'{o}lica de Chile, Departamento de Astronom\'{\i}a y
  Astrof\'{\i}sica, Casilla 306, Santiago 22, Chile\\
  \email{cccacere@astro.puc.cl}
  \and
  European Southern Observatory, Av. Alonso
  de Cordova 3107, Santiago 19001, Chile
  \and
  Specola Vaticana, V00120 Vatican City State, Italy.
  \and 
  Departamento of Astronom\'{\i}a, Universidad de Concepci\'{o}n,
  Casilla 160-C, Concepci\'{o}n, Chile.\\
}

\date{Received ??? ??, ????; accepted ??? ??, ????}

\abstract {Currently the only technique sensitive to Earth mass
  planets around nearby stars (that are too close for microlensing) is
  the monitoring of the transit time variations of the transiting
  extrasolar planets. We search for additional planets in the systems
  of the hot Neptune GJ\,436b, and the hot-Jupiter XO-1b, using
  high cadence observations in the $J$ and $K_S$ bands. New
  high-precision transit timing measurements are reported: GJ\,436b
  $T_C = 2454238.47898 \pm 0.00046$ HJD; XO-1b $T_C(A) =
    2454218.83331\pm0.00114$ HJD, $T_C(B) = 2454222.77539\pm 0.00036$
    HJD, $T_C(C) = 2454222.77597 \pm 0.00039$ HJD, $T_C(D) =
    2454226.71769\pm0.00034$ HJD, and they were used to derive new
    ephemeris. We also determined depths for these transits. No
  statistically significant timing deviations were detected. We
  demonstrate that the high cadence ground based near-infrared
  observations are successful in constraining the mean transit time to
  $\sim 30$\,sec., and are a viable alternative to space missions.}

\keywords{Stars: planetary systems -- Stars: individual: GJ\,436, XO-1 --
  Methods: observational}

\maketitle
%

\section{Introduction}
More than 300 extrasolar planets are known up to date. Most of them
were discovered by radial velocity searches. At this time nearly sixty
transiting systems have been detected. Some of them reside in crowded
fields, making them difficult targets for follow up studies. The small
amplitude of the transits ($\leq$3\%) and the low probability of the
suitable geometric configurations (orbital plane along the line of
sight, i.e. edge-on orbit) partly explain the number of detections.

Assuming that the host star radius is known, transits allow a direct
determination of the orbital plane inclination and the planet
radius. When combined with radial-velocity data, they give access to
the real planet mass and to the mean planet density revealing its
nature: gaseous, icy or rocky. Transits are potentially sensitive to
Earth size planets which cause too small movement of the host star to
be detected via radial velocity. Of course, direct occultations of
stars by Earth like planets are also difficult to detect and obtaining
adequate observations is feasible only with space missions.

A numerical investigation of the transit timing sensitivity to
perturbing planets was reported in \citet{agol_etal05}, and in
\citet{holman_and05}. The transit time variations (TTVs) depend on the
detailed configuration of the system, but some fairly robust general
predictions are possible for the case of hot Jupiters (such as most of
the transiting planets known so far): (i) TTVs are largest in case of
resonant orbits; (ii) TTVs are proportional to the mass of the
perturbing planet; (iii) TTVs are proportional to the period of the
transiting planet; (iv) TTVs vary with time \citep[see Fig.\,6
in][]{agol_etal05}, and in an optimal orbital configuration the
interval between two sequential transits can change by as much as few
minutes. For example, an Earth mass planet in the system of
HD\,209458, in a 2:1 resonance with HD\,209458b will lead to TTV of
order of 3 min.

The probability for observing long period transiting planets is low
and all known transiting planets have periods $\leq$9.2\,days, with
the only exceptions of the high-eccentricity planets HD\,17156b
\citep{barbieri_etal07}, and HD\,80606b \citep{moutou_etal09}, with
orbital periods of $\sim21d$, and $\sim111d$ respectively. However,
systems of multiple extrasolar planets on resonant orbits were found
by the radial velocity searches (i.e. 51 Cancri, 3:1 resonance;
HD\,82943 and Gl\,876 -- 2:1), showing that this is a realistic
possibility. The available data show that at least 12\% of known
extrasolar planetary systems have more than one planet, and possibly
the real fraction is much larger.

The insufficient timing accuracy of individual transits limited the
planet searches with this method so far: most timing data came from
small telescopes with large ``dead'' time between the images, in some
cases 2/3 or more of the observation duty-cycle. Furthermore, many of
these observations require defocusing of the telescope (usually
because of the large pixel sizes) leading to contamination from
fainter neighboring stars, sometimes equal to the flux of the host
star \citep{bakos_etal06}.

The first detailed TTV study \citep{steffen_and05} used the
observation of 12 transits of TrES-1b by \citet{alonso_etal04} and
\citet{charbonneau_etal05} to search for additional planets in this
system. They found no convincing evidence for a second planet, and
they can only set an upper mass limit for planets in low order
resonances comparable or lower than the Earth mass,~making these
timing data the first that are sensitive to Earth mass perturbing
planets.

The possibility of having another planet in the system of OGLE-TR-113b
was studied by \citet{gillon_etal06} -- their maximum TTV
  amplitude is 43\,sec (2.5$\sigma$), consistent with $\le$\,1-7
Earth mass planets. \citet{bakos_etal06} derived residuals for
individual transits for HD\,189733b ranging from 0.7\,sec to 302\,sec
or 0.0045-5.9$\sigma$, and the errors of individual transit times are
19-150\,sec. They are consistent with perturbations from
0.15\,M$_{\mathrm {Jup}}$ mass planet at 2:1 resonance orbit that
would remain undetected in radial velocity observations.  The authors
refrain from strong statement because the data are affected by
systematic errors. Similarly, \citet{diaz_etal08} found variations in
the period of transits of the planet OGLE-TR-111 whose origin has not
been conclusively determined.

Space based photometry developed by the MOST team has provided
accurate values for the transit times on the HD\,189733 and HD\,209458
systems \citep{miller-ricci_etal08a, miller-ricci_etal08b}. These
results have ruled out the presence of super-Earths in the inner
resonances. Analysis of the transit times for ground-based
observations of various transiting systems have been performed by the
Transit Light Curve (TLC) team \citep[e.g.][]{holman_etal06,
  winn_etal07}, who have obtained accurate timing values which have
show no strong evidence for the presence of a third body in the
systems.

Here we describe the first results from our timing study of individual
transits with infrared (IR) detectors that allow us to obtain imaging
with minimum ``dead'' time for readout ($\leq$0.1\%). By design the IR
detectors read out faster than the CCDs because the high background
forces the usage of short exposures and the IR array technology has
evolved to achieve reset/readout times of order of microseconds,
rather than the many seconds needed to shift the charges across the
CCDs. This gives us the following advantages: (i) we observed with
unprecedented time resolution of $\sim$0.1-0.2\,sec; the host stars on
individual images have S/N$\sim$50-100, depending on the band and the
target brightness; (ii) we observed bright planet hosting stars
without defocusing (the exception was one transit observation of
GJ-436b), as it is often the case with the previous studies that
attempted to use larger telescopes, reducing the contamination from
nearby sources; (iii) we relied on the ESO timing system that provided
us with uniform time accurate to better than 0.1\,sec, minimizing any
systematic effects -- a crucial advantage over other studies that rely
on collecting data from various telescopes.  Our preliminary
simulations that included only well-behaved Poisson photon noise
suggested a transit timing accuracy of 0.1-1\,sec
\citep{ivanov_etal09}.

We apply this high-cadence method to GJ\,436b -- the first
transiting hot Neptune planet reported \citep{gillon_etal07a}, hosted
by a M2.5V star. The small size of GJ\,436b leads to a challengingly
shallow depth of only 0.6\%. This planet is particularly interesting
because it shows large eccentricity which may be caused by the
gravitational perturbation of a third body in the system. The possible
presence of a super-Earth near a 2:1 mean motion resonance was
proposed by \citet{ribas_etal08}. This scenario was recently ruled
out, but the presence of a third body may still be possible
\citep{ribas_etal09}. We also apply this method to the Jupiter-mass
planet XO-1b, hosted by a Sun-like star \citep{mccullough_etal06}, in
a $\sim$\,4\,d orbit.

\section{Observations and Data Reduction}
The XO-1b data were collected with the SofI (Son of ISAAC) instrument
at the 3.6-m ESO New Technology Telescope (NTT) on La Silla,
and with the ISAAC (Infrared Spectrometer And Array Camera) instrument
at the 8.2-m UT1 (Antu) unit of the ESO Very Large Telescope
on Cerro Paranal, in Visitor Mode, and GJ\,436b data were collected
only with SofI. All observations were carried out in the {\sl
  Fast-Phot} cube mode, which produces a series of data-cubes with
short integration times, and with virtually zero dead time between
integrations, because of the reduced communications between the
detector and the instrument workstation in this mode. During the
observations the detector was windowed down to minimize the readout
and data transfer overheads, with the requirement that the field of
view contained the target and a reference star of a similar
brightness, which was used for differential photometry. The windowing
allows us to reduce the detector integration time to less than
0.01\,sec, if necessary. The typical overhead is $\sim6$\,sec per cube
of 100-2000 frames. A summary of the observing details is presented in
Table~\ref{table:0}.

\begin{table*}
\caption{Observations and data reduction summary.}              
\label{table:0}      
\renewcommand{\footnoterule}{}  
\begin{minipage}[t]{2\columnwidth}
\begin{tabular}{c@{ }c@{ }c@{ }c@{ }c@{ }c@{ }c@{  }c@{  }c@{\hspace{4pt}}c}
\hline \hline 
 Target & Run & Date & Instrument & Window size (px) & DIT (sec) & N.\,Frames & Filter &
Apertures (px) & Inner annuli (px)\\
\hline
GJ\,436b & A & May 17, 2007 & SofI & $872\times132$ & 0.239 & 36\,058 &
$K_S$ & 23 & 35\\

XO-1b & A & April 27, 2007 & SofI & $352\times152$ & 0.8 & 17\,066 &
$J$ & 11 & 19\\

XO-1b & B & May 1, 2007 & SofI & $892\times180$ & 0.8 & 20\,100 & $J$ &
8 & 20\\

XO-1b & C & May 1, 2007 & ISAAC & $64\times544$ &
0.17 - 0.13\footnote{The DIT (Detector Integration Time) was reduced during the observations to
  avoid reaching the non-linear limit of the detector in improving sky
  conditions.} & 110\,161 & $K_S$ & 7 & 10 \\

XO-1b & D & May 5, 2007 & ISAAC & $64\times544$ & 0.08 & 186\,095 & $J$+Block &
21 & 21\\
\hline
\end{tabular}
\end{minipage}
\end{table*}

\subsection{SofI Observations}
SofI is the infrared camera and spectrograph at the NTT telescope on
La Silla \citep{moorwood_etal98a}. It is equipped with a Hawaii HgCdTe
array of 1024x1024 pixels, with a gain of 5.4\,e$^-$\,ADU$^{-1}$ and a
readout noise of 2.1\,ADU. Its detector shows a non-linearity of less
than 1.5\% below 10,000\,ADU, in a Correlated Double Sampling
readout. The imaging mode has a pixel scale of
0.288\,arcsec\,pix$^{-1}$.

The observations of GJ\,436b were carried out the night of May 17,
2007, in poor weather conditions. We applied a linearity
correction to the data based on calibration data obtained on May 14,
2007, where a 4-order polynomial was fitted to the deviation from
linear detector response\footnote{More information can be found at the
  SofI web page
  http://www.eso.org/sci/facilities/lasilla/instruments/sofi/inst/Linearity.html}.
Despite of defocusing the telescope in this run, some images show
pixels values above the correctable 18\,000 counts level, and they
were omitted from the resulting light curve, such that the final
sample spans 35\,266 points, covering $\sim 146$ minutes.

We observed two transits of XO-1b with SofI, during the nights of
April 27 (run A), and May 1 (run B), 2007, both in the $J$ band,
covering 228\,min and 303\,min, respectively.

\subsection{ISAAC Observations}
ISAAC is an infrared camera and spectrograph located at the Nasmyth B
focus of UT1 \citep{moorwood_etal98b}. For our observations we used
the long-wavelength arm, which equipped a 1024x1024 Aladdin array,
with a pixel scale of 0.148\,arcsec\,pix$^{-1}$, a gain of
8.7\,e$^-$\,ADU$^{-1}$, and a read noise of 4.6 ADU. The readout mode
was Double Correlated Read Low Bias. This detector is linear at 90\%
for signal below 16,000\,ADU.

We observed two transits of XO-1b, during the nights of May 1 (run C),
2007, in the $K_S$ band, and May 5 (run D), 2007, in the $J$ band,
using the ISAAC $J$+Block filter\footnote{The $J$ filter suffers from
  a red leak.  Normally, it is eliminated by the sensitivity cut-off
  of the Hawaii detectors at $\sim 2.5\mu m$, but the Aladdin detector
  used in the long-wavelength arm of ISAAC is sensitive to $\sim 5\mu
  m$. A blocking filter is added to eliminate the leak. The overall
  transmission of the $J$+Block filter is similar to that of the
  ``standard'' $J$ filter.}. These observations cover 277\,min and
257\,min, respectively.

\subsection{Data Reduction}
Standard infrared data reduction steps were applied: flat fielding,
and dark subtraction. However, since the observations were obtained in
stare mode, i.e. with no jittering, the sky was not subtracted with
the usual method used for infrared data. Instead, we estimated the sky
level by measuring the flux in circular annuli centered on the target
and the reference star. Note that PSF fitting was not possible because
of the defocusing of the GJ\,436b observations and even if it would
not have been the case, the targets are the brightest sources in the
field, and the seeing variation did not allow to create a PSF model
from the previous and/or next frames. The fundamental limit of
  how bright the reference star could be comes from the maximum size
  of the detector window. We always select as a reference source the
  brightest available star in the field of view, and if it is fainter
  than the target, it dominates the noise of the final light curve, as
  in the case of GJ\,436b. The data reduction was carried out with the
IRAF\footnote{IRAF is distributed by the National Optical Astronomy
  Observatories, which are operated by the Association of Universities
  for Research in Astronomy, Inc., under cooperative agreement with
  the National Science Foundation.}  package DAOPHOT.

The final light curves were divided by a linear polynomial of the form
$Correction = a + b\times T$, where $T$ is the time, calculated with
the out-of-transit points of each light curve, to normalize the light
curve, and correct any smooth trend due to atmospheric variations.

The five light curves obtained are presented in Table
\ref{table:photometry}, where we only show a small fraction of the
complete light curves as a format example. The complete light curves
are available in the electronic version of the Journal.

\begin{table}
\begin{minipage}[t]{0.8\columnwidth}
\caption{Transit light curves.}              
\label{table:photometry}      
\renewcommand{\footnoterule}{}  
\begin{tabular}{c c c}          
\hline
\hline                        
HJD & Flux & Error\\    
\hline                                   
\multicolumn{3}{c}{}\\  
\multicolumn{3}{c}{GJ 436b}\\
2454238.456329 & 0.986084 & 0.005644\\
2454238.456332 & 0.995099 & 0.005565\\
2454238.456335 & 1.005212 & 0.005784\\
2454238.456337 & 1.018050 & 0.005828\\
2454238.456340 & 1.015619 & 0.005785\\
\multicolumn{3}{c}{...}\\
\multicolumn{3}{c}{}\\  
\multicolumn{3}{c}{XO-1b Run A}\\
2454218.911636 & 1.003098 & 0.003059\\
2454218.911645 & 1.017395 & 0.003063\\
2454218.911654 & 1.017412 & 0.003075\\
2454218.911663 & 0.991108 & 0.002969\\
2454218.911673 & 0.988897 & 0.003064\\
\multicolumn{3}{c}{...}\\
\multicolumn{3}{c}{}\\  
\hline                                             
\end{tabular}
\end{minipage}
\end{table}

\section{Analysis}
We focus on measuring the central transit time ($T_c$) and depth
($d$). For GJ\,436b we adopted the stellar parameters determined by
\citet{gillon_etal07a}: $T_{eff} = 3500$\,K, $\log g = 4.5$, and
[Fe/H] = 0.0. The rest of the system parameters were taken from
\citet{torres07}.  For XO-1b, we adopted stellar parameters from
\citet{mccullough_etal06}: $T_{eff} = 5750$\,K, $\log g = 4.53$, and
[M/H] = 0.058. Planetary and orbital parameters were taken from
\citet{holman_etal06}.  The transit length was calculated from the
known orbital parameters and the depth was measured as ratio of the
out-of-transit to the in-transit flux.  The next step was to create a
light curve model according to the prescription of
\citet{mandel_and02}, assuming throughout the paper a quadratic
limb-darkening law, where the limb-darkening coefficients were taken
from \citet{claret00} for the adopted stellar parameters, and the used
band passes.

The final step was to fit this light curve to the observations
minimizing the $\chi^2$ statistics:
\begin{equation}
\label{eq:chi2}
\chi^2 = \sum_{i=1}^{N} \left[ \frac{f_i -
    f_i^O}{\sigma_i^O}\right]^2
\end{equation}
where $f_i^O$ is the observed flux with an uncertainty $\sigma_i^O$,
and $f_i$ is the expected flux obtained from the model. Here, the
central transit time ($T_C$) was the only free parameter. The
minimization method of \citet{brent73}, as implemented in
\citet{press_etal92}, was used.

\subsection{GJ\,436b}
The transit depth we measured for GJ\,436b is $d = 0.64 \pm
0.03\%$. This corresponds to a planet-to-star size ratio: $p = 0.082
\pm 0.002$ which is in good agreement with the \citet[][$p=0.0829 \pm
0.0043$ and $p=0.082 \pm 0.005$, respectively]{gillon_etal07b,
  gillon_etal07a}, and marginally with \citet[][$p=0.0839 \pm
0.0005$]{deming_etal07}.  For the values given above, we adopted the
corresponding limb-darkening coefficients in $K$\footnote{We assume
  the same limb darkening coefficients for $K$ and $K_S$ because the
  effective wavelengths of the two filters are similar.}: $a=-0.0677$,
$b=0.3665$.

\begin{figure}
  \centering
  \includegraphics[scale=0.45]{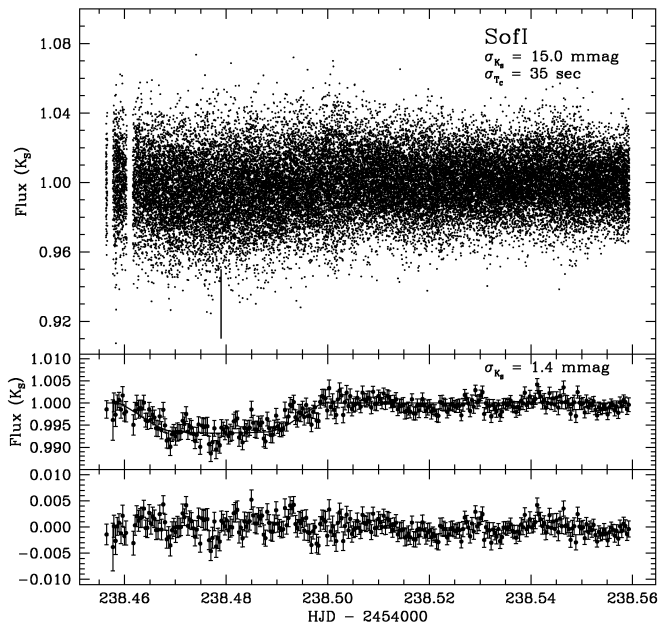}
  \caption{Photometry of the GJ\,436b transit obtained with SofI. {\it
      Top panel}: the normalized light curve, with a integration time
    of 0.239\,sec. The central time of the transit is marked at HJD
    $2454238.47898$. Error-bars are omitted for the sake of
    clarity. The out-of-transit r.m.s. flux on the unbinned curve
    ($\sigma_{K_S}$), and the timing error ($\sigma_{T_C}$) are also
    shown. {\it Middle panel}: the binned curve, and the best fitting
    model. The bin width is 30\,sec. Error bars represent the
    flux r.m.s. within each bin. Note the different Y-axis range on
    the panels. The out-of-transit r.m.s. flux ($\sigma_{K_S}$) of
    the binned curve is also presented. {\it Bottom panel}: The residuals of the best
    fitting model to the binned curve.}
  \label{fig:gj-436b}%
\end{figure}

We fitted the model to the observed light curve to obtain the transit
midpoint $T_C = 2454238.47898 \pm 0.00046$\,HJD ($\sim 39$~sec error).
The observed light curve for GJ\,436b is shown in
Fig. \ref{fig:gj-436b} (top panel), together with a light curve binned
to 30\,sec time resolution, and the fitting model (middle panel), and
the residuals of the binned curve (bottom panel). The binned light
curve was used in this analysis. The time and the flux values of a bin
are the average of the times and fluxes of all measurements within the
bin, and the flux error is their r.m.s.

We performed a bootstrapping simulation to calculate the time of
transit uncertainty. The set of residuals of the best-fitting model
was shifted a random number of points in a circular way, and then
added to the model light curve, constructing a simulated light curve
with the same point-to-point correlation as the observed light
curve. This procedure takes into account the correlated noise in our
analysis. Then, we calculated the center-of-transit for the new curve,
as described above. This procedure was repeated 10000 times, and the
1-$\sigma$ width of the resulting distribution of timing measurements
was adopted as the error of the timing. The 1-$\sigma$ error value is
weakly dependent on binning. Here we choose a bin size of 30\,sec
because it is a good compromise between the number of points included
in each bin and the final number of points in the light curve.

\begin{figure}
  \centering
  \includegraphics[scale=0.45]{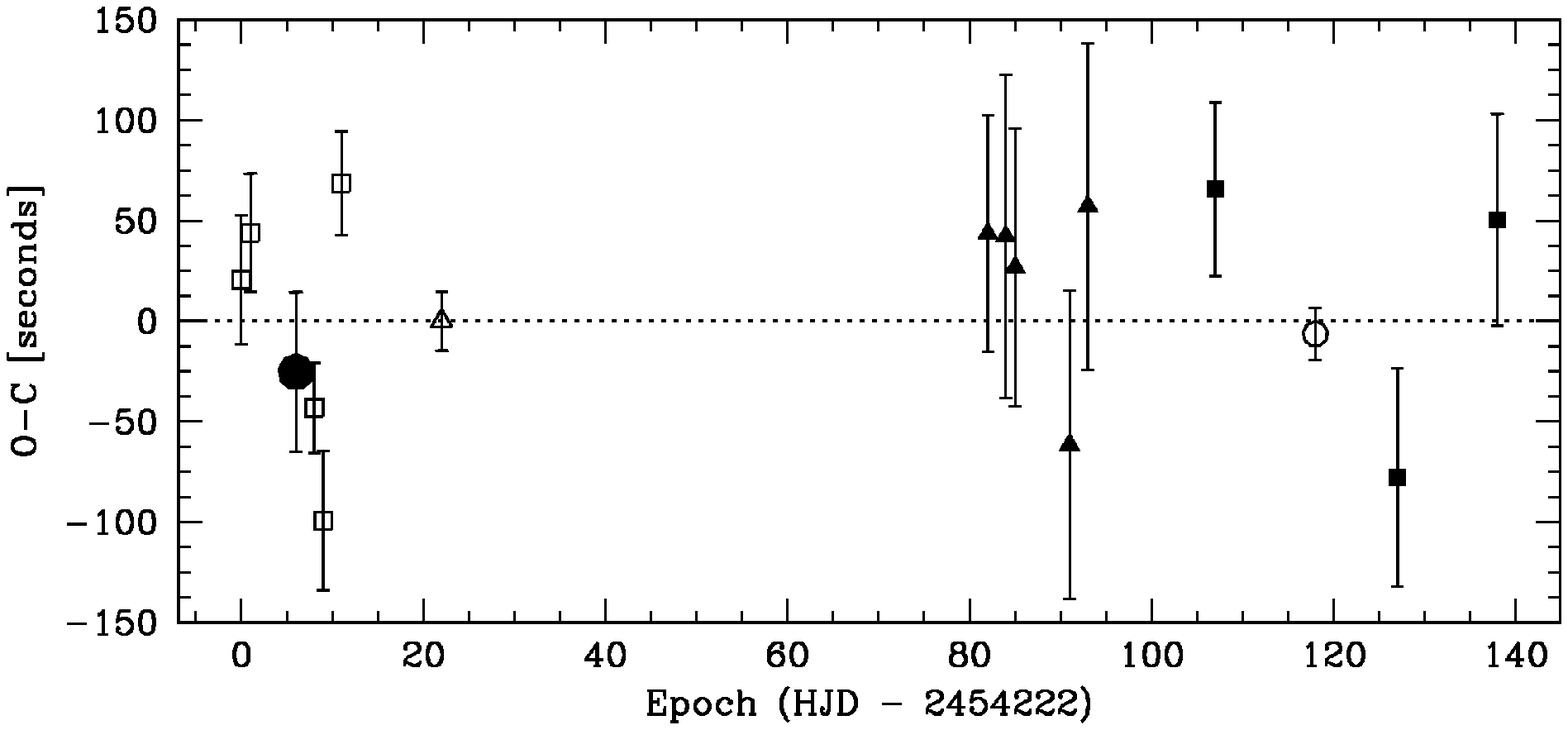}
  \caption{The observed minus calculated time-of-transit diagram for
    the different timing values in the literature, as a function of
    the observed epoch, for GJ\,436b. Different symbols represent
    different literature sources. The open squares are from
    \citet{shporer_etal08}. The open triangle is an average of the
    value from \citet{gillon_etal07b}, \citet{deming_etal07}, and
    \citet{southworth08}, with the corrections given by
    \citet{bean_etal08}. The solid triangles are from
    \citet{bean_and08}. The open circle is from
    \citet{alonso_etal08}. The solid squares come from
    \citet{ribas_etal09}. Finally, our timing estimate is drawn with a
    solid circle at $E=6$. All measurements are listed in Table
    \ref{table:1}}
  \label{fig:o-c}%
\end{figure}

Many follow-up observations of GJ\,436b has been carried out with both
\emph{Spitzer Space Telescope} \citep{gillon_etal07b, deming_etal07,
  demory_etal07, southworth08} and the \emph{Hubble Space Telescope}
\citep{bean_and08}, and recently ground based observations have given
a timing precision comparable to space based observations
\citep{alonso_etal08, shporer_etal08}. They are all listed in Table
\ref{table:1}.

Considering the newest literature data, and our measurement we
recalculated the ephemeris of GJ\,436b by fitting a weighted linear
relation, to obtain a period $P=2.6438986 \pm 0.0000016$\,d, and
a ``zero transit'' epoch $T_{C}(E=0) = 2454222.61588 \pm 0.00012$
  HJD. The new epoch presented in this work is in excellent agreement
with these ephemeris, and they in turn agree with those of
  \citet{ribas_etal09}, and \citet{bean_and08}. Figure \ref{fig:o-c}
shows the residuals of the fit of the new ephemeris as a function of
the observed epoch for the available timing values in the literature,
and our timing value at epoch $E=6$.

The data show some TTVs of up to 98\,sec. -- smaller than
the predicted deviations of order of a few minutes for a 1-10 Earth
mass companion on a resonant 2:1 orbit \citep{alonso_etal08}. However,
this deviations are consistent with zero, within their respective
uncertainties. Further observations with higher accuracy are necessary
to constrain better the properties of this system and to address the
question if it bears other planets.

\begin{table}
\begin{minipage}[t]{0.8\columnwidth}
\caption{GJ\,436b timing measurements.}              
\label{table:1}      
\renewcommand{\footnoterule}{}  
\begin{tabular}{c c c c}          
\hline
\hline                        
$T_C$ (HJD) & $\sigma_{T_C}$ (d) & Epoch & Reference \\    
\hline                                   

2454222.61612       & 0.00037 & 0   & (1)\\
2454225.26029       & 0.00038 & 1   & (1)\footnote{Average of the two values given
  for this epoch.}\\
{\bf 2454238.47898} & {\bf 0.00040} & {\bf 6} & {\bf This work}\\
2454243.76657       & 0.00026 & 8   & (1)\\
2454246.40982       & 0.00040 & 9   & (1)\\
2454251.69956       & 0.00030 & 11  & (1)\\
2454280.78165       & 0.00017 & 22  & (2,3,4)\footnote{Average of the three values given
  by different authors for this epoch, considering the correction
  given by Bean et al. (2008)}\\
2454439.41607       & 0.00068 & 82  & (5)\\
2454444.70385       & 0.00093 & 84  & (5)\\
2454447.34757       & 0.00080 & 85  & (5)\\
2454463.20994       & 0.00089 & 91  & (5)\\
2454468.49911       & 0.00094 & 93  & (5)\\
2454505.51379       & 0.00050 & 107 & (6)\\
2454534.59584       & 0.00015 & 118 & (7)\\
2454558.39010       & 0.00063 & 127 & (6)\\
2454587.47447       & 0.00061 & 138 & (6)\\
\hline                                             
\end{tabular}
References: (1) - \citet{shporer_etal08}; (2) -
\citet{gillon_etal07b}; (3) - \citet{deming_etal07}; (4) -
\citet{southworth08}; (5) - \citet{bean_and08}; (6) -
\citet{ribas_etal09}; (7) - \citet{alonso_etal08}.
\end{minipage}
\end{table}

\begin{figure*}
  \begin{center}
    \includegraphics[width=8.7cm,height=9.7cm]{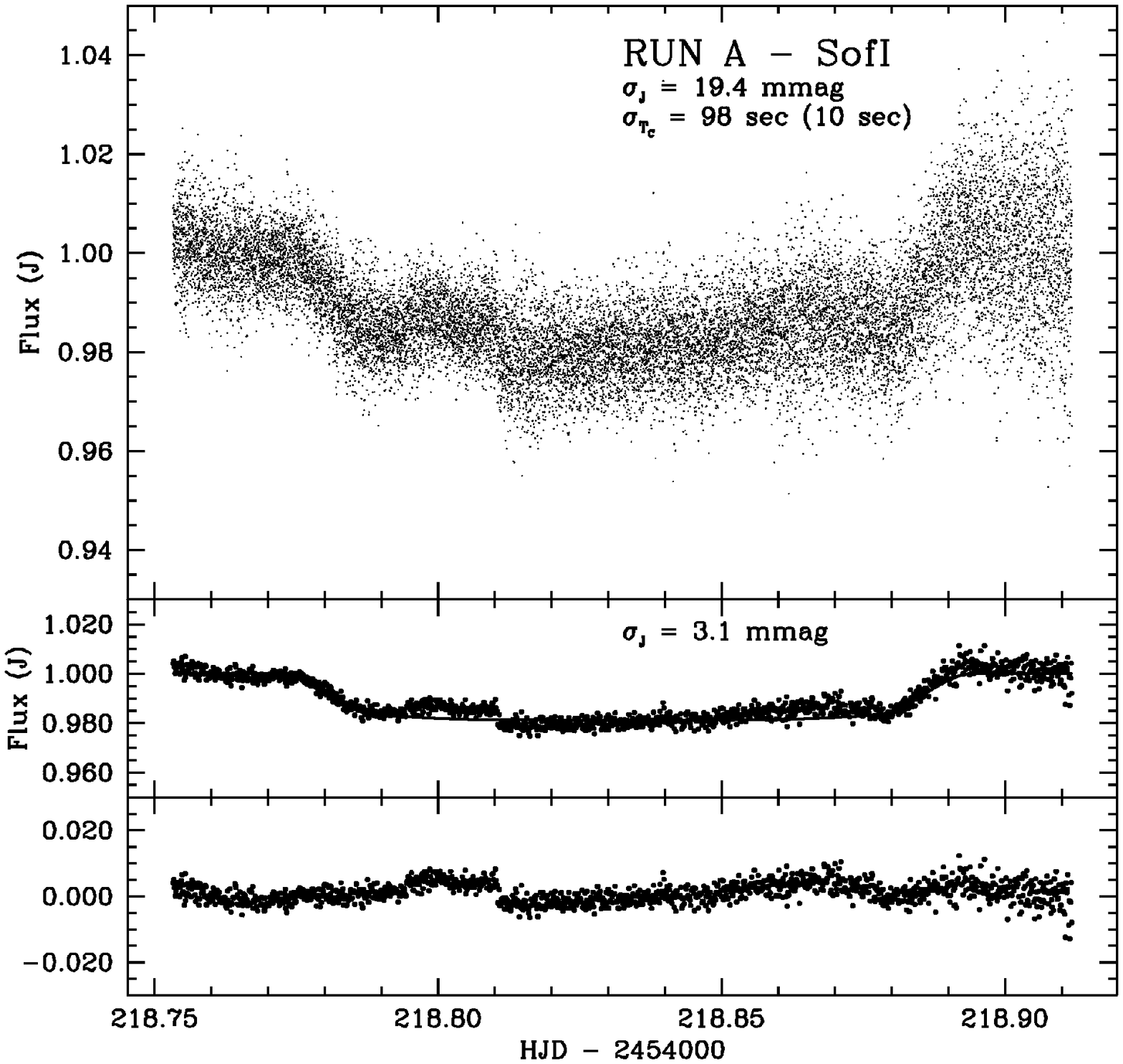}
    \includegraphics[width=8.7cm,height=9.7cm]{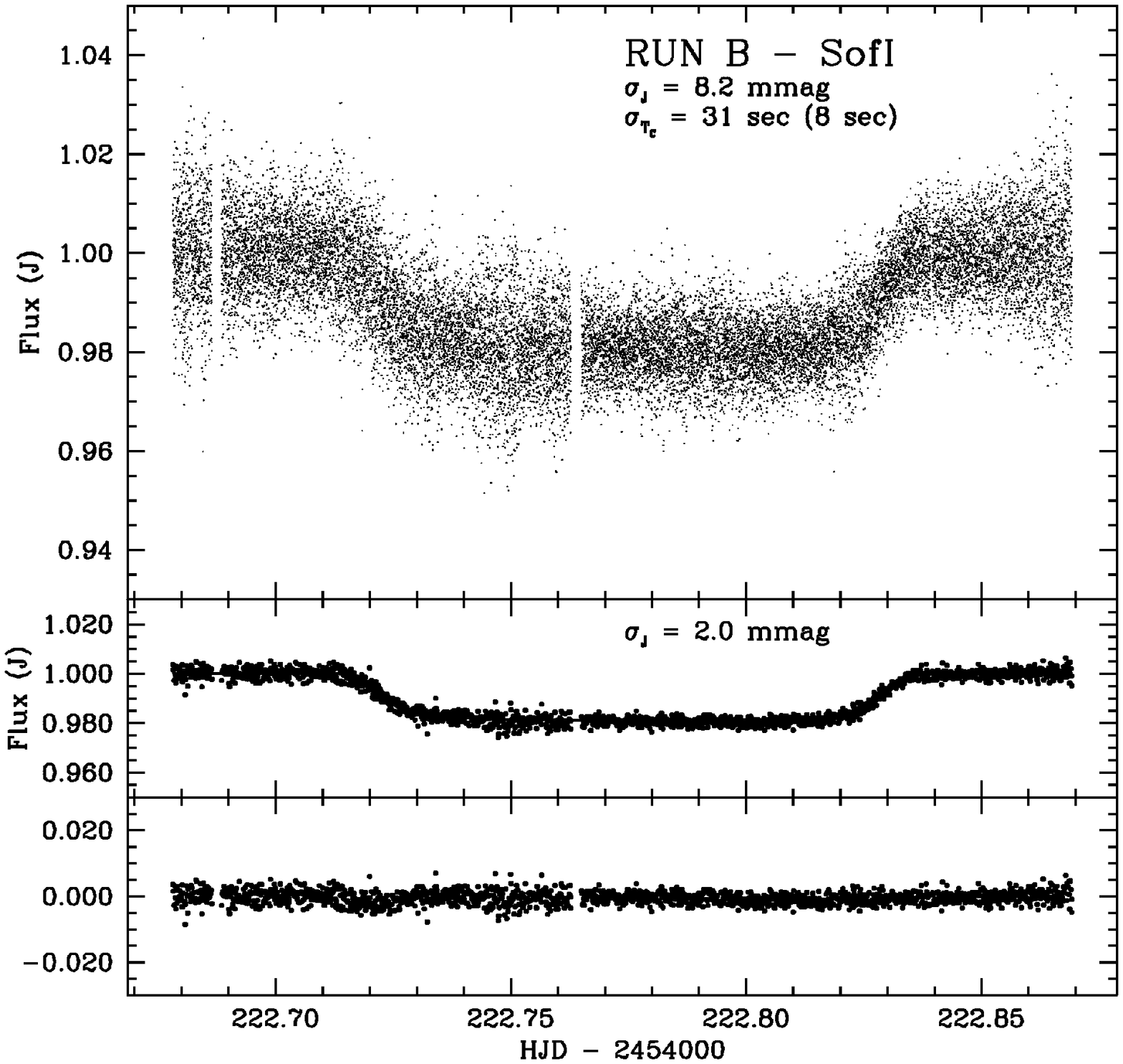}\\
    \includegraphics[width=8.7cm,height=9.7cm]{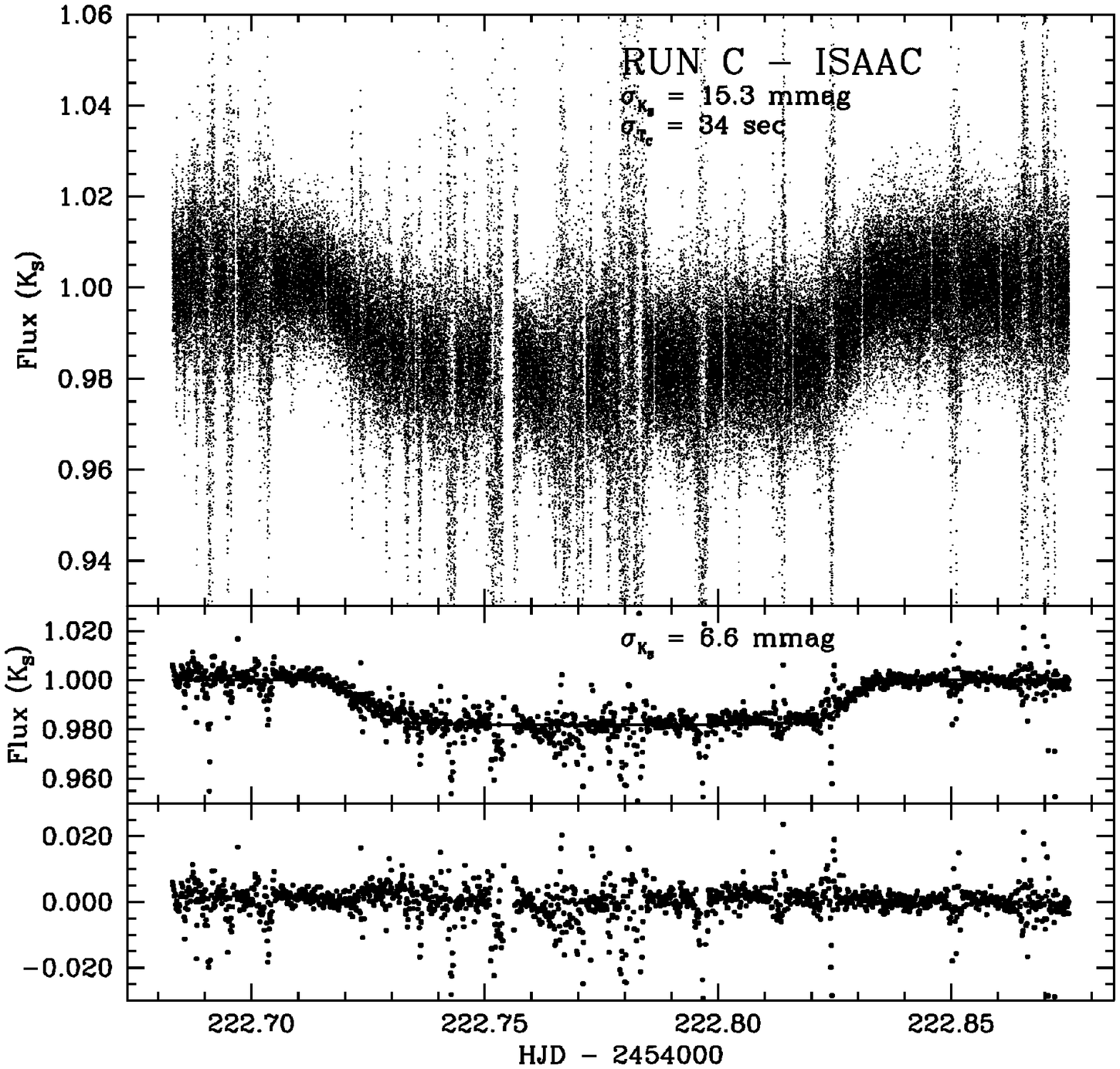}
    \includegraphics[width=8.7cm,height=9.7cm]{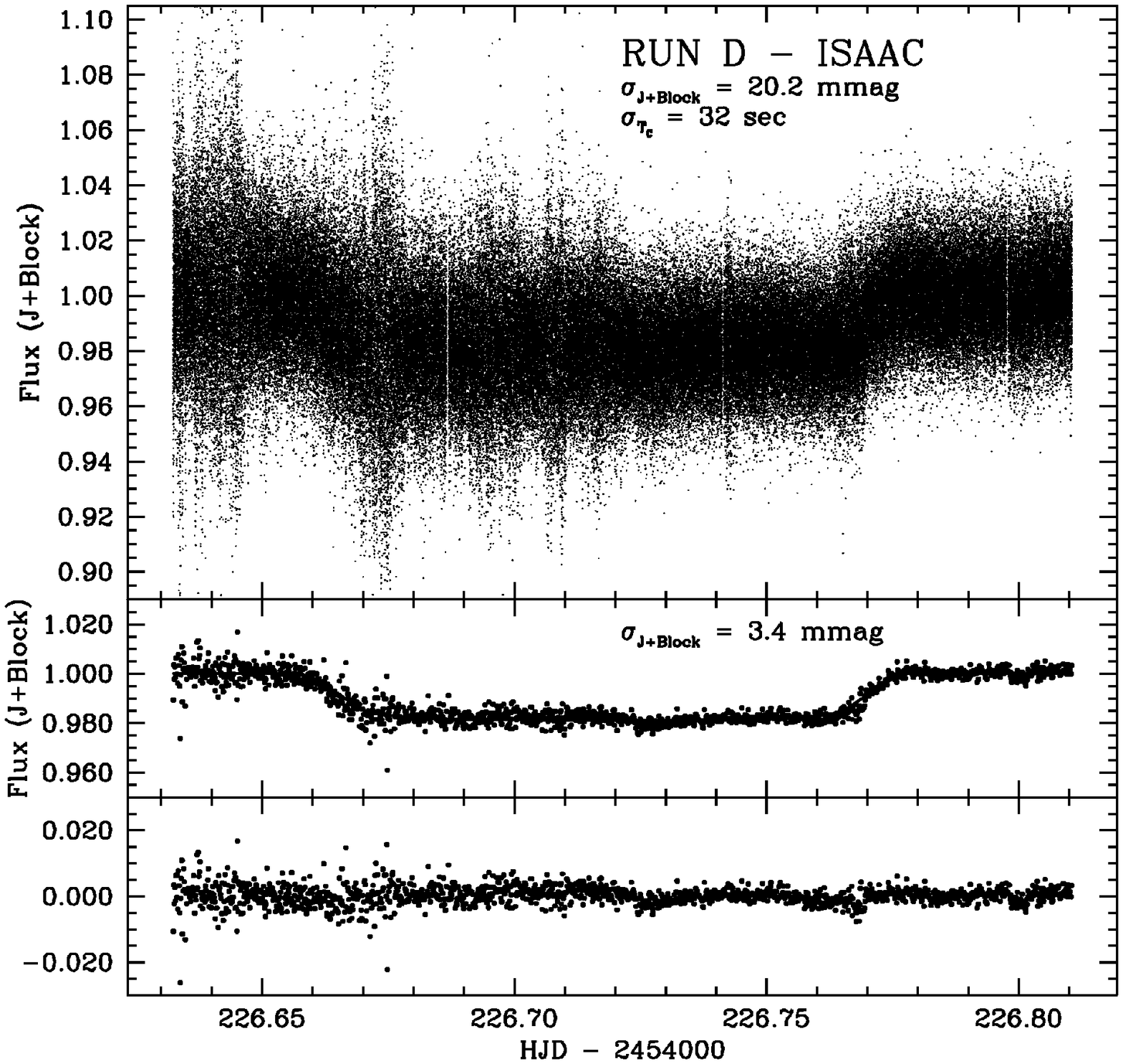}
    \caption{Photometry of the XO-1b transits obtained with SofI and
      ISAAC. Each panel shows (from top to bottom) the original light
      curve with a 10\,sec bin size version with the best-fitting
      model, and the residuals of the fit. {\it Top-left panel}: Run
      A, observed with SofI, in the $J$ band. {\it Top-right panel}:
      Run B, observed with SofI, in the $J$ band. {\it Bottom-left
        panel}: Run C, observed with ISAAC, in the $K_S$ band. {\it
        Bottom-right panel}: Run D, observed with ISAAC, in the
      $J$+Block band. Each panel gives the out-of-transit r.m.s. flux
      of the unbinned curve (upper sub-panel), and of the binned curve
      (middle sub-panel). The timing error ($\sigma_{T_C}$) for each
      transit is printed. Note that for runs A and B we also give, in
      brackets, the timing error obtained if only the ingress-egress
      phases are considered. See more details in
      Sect. \ref{section:analysis_xo-1b}}
  \label{fig:xo-1b}%
  \end{center}
\end{figure*}

\subsection{XO-1b}
\label{section:analysis_xo-1b}

The XO-1b light curves were analyzed in a similar way as for
GJ\,436b. The quadratic stellar limb-darkening coefficients utilized
here were: $a=0.00592$ and $b=0.34954$ for the $K_S$ light curve, and
$a=0.10923$ and $b=0.35938$ for the $J$+Block light curve.

After fixing the system parameters we calculated the transit midpoint
for the 4 light curves separately. The errors were calculated with the
same bootstrapping technique described above. The transit timings
calculated here are shown in Table \ref{table:2}.

Interestingly, the uncertainties of the runs A and B decrease
significantly, if the $\chi^2$ is calculated only over the ingress and
the egress phases: 0.00012\,d and 0.00009\,d (10 and 8\,sec),
respectively. The bootstrapping simulation was also done over these
parts of the light curve.  Apparently, using only the ingress and
egress excludes some of the systematic effects that occurred during
the rest of the transit, and were reflected in the error
distribution. Therefore, we consider the errors given in Table
\ref{table:2} to be upper limits of the uncertainties. The errors of
the Runs C and D remained virtually unchanged: 0.00044, in both cases,
most likely because they were obtained in poor weather conditions and
the errors are dominated by the reduction of flux during the periods
of poor atmospheric transmission.

\begin{table}
\begin{minipage}[t]{0.8\columnwidth}
\caption{XO-1b timing measurements.}              
\label{table:2}      
\renewcommand{\footnoterule}{}  
\begin{tabular}{c c c c}          
\hline
\hline                        
$T_C$ (HJD) & $\sigma_{T_C}$ (d) & Epoch & Reference \\    
\hline                                   
2453127.03850  & 0.00580 & -173 & (1)\\
2453142.78180  & 0.02180 & -169 & (1)\\
2453150.68550  & 0.01060 & -167 & (1)\\
2453154.62500  & 0.00260 & -166 & (1)\\
2453158.56630  & 0.00340 & -165 & (1)\\
2453162.51370  & 0.00250 & -164 & (1)\\
2453166.45050  & 0.00250 & -163 & (1)\\
2453170.39170  & 0.00370 & -162 & (1)\\
2453229.51430  & 0.00450 & -147 & (1)\\
2453237.40430  & 0.00320 & -145 & (1)\\
2453241.34100  & 0.00670 & -144 & (1)\\
2453808.91700  & 0.00110 &    0 & (2)\\
2453875.92305  & 0.00036 &   17 & (3)\\
2453879.86400  & 0.00110 &   18 & (3)\\
2453883.80565  & 0.00019 &   19 & (3)\\
2453887.74679  & 0.00016 &   20 & (3)\\
{\bf 2454218.83331}  & {\bf 0.00114} &  {\bf 104} & {\bf This work (A)}\\
{\bf 2454222.77539}  & {\bf 0.00036} &  {\bf 105} & {\bf This work (B)}\\
{\bf 2454222.77597}  & {\bf 0.00039} &  {\bf 105} & {\bf This work (C)}\\
{\bf 2454226.71769}  & {\bf 0.00037} &  {\bf 106} & {\bf This work (D)}\\
\hline                                             
\end{tabular}
References: (1) - \citet{wilson_etal06}; (2) - \citet{mccullough_etal06}; (3) - \citet{holman_etal06}.
\end{minipage}
\end{table}

The final light curves are shown in Fig. \ref{fig:xo-1b}, with a
20\,sec bin width version to easily show the best fitting model. Run A
shows some systematics that could not be corrected, so this curve was
only used to get timing value, and not planetary parameters.

\begin{figure}
  \centering
  \includegraphics[scale=0.48]{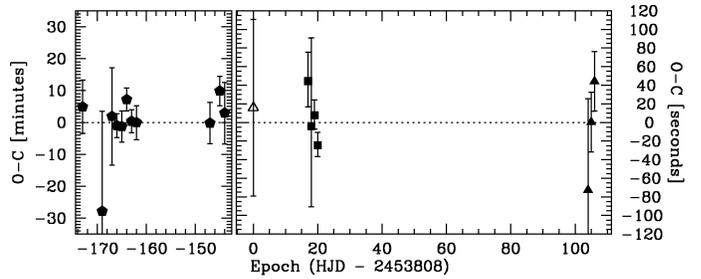}
  \caption{The observed minus calculated time-of-transit diagram for
    the different timing values in the literature, as a function of
    the observed epoch for XO-1b. For clarity the diagram was divided
    in two different scale panels (minutes in the left panel, and
    seconds in the right panel). Different symbols represents
    different literature sources. The pentagons are from
    \citet{wilson_etal06}. The squares are the values from
    \citet{holman_etal06}. The open triangle is the value from
    \citet{mccullough_etal06}. Our timing values are shown with solid
    triangles. All measurements are listed in Table \ref{table:2}.}
  \label{fig:o-c_xo-1b}%
\end{figure}

We fitted a weighted linear relation to the timing values listed in
Table \ref{table:2}, to get the predicted ephemeris for the transits
of XO-1b. This new ephemeris correct the long term difference in the
ephemeris given by \citet{mccullough_etal06} and
\citet{wilson_etal06}. Our fit gives us: $P=3.9415128 \pm
0.0000028$\,d, and a ``zero transit'' epoch $T_{C}(E=0) =
2453808.91682 \pm 0.00013$ HJD. In this calculation, we use the
weighted average of runs B, and C, which spans the same epoch. This
calculations are shown in Fig. \ref{fig:o-c_xo-1b}. Note that runs C
and D, carried out with the bigger telescope, but under worse weather
conditions yield less accurate timing measurements than transits
observed with the smaller telescope but under better weather
conditions (i.e. run B), demonstrating the impact of the weather on
the high precision photometry required to detect the transiting
planets.

The amplitude of the resulting TTVs, show no evidence for a perturbing
third body in the system, in agreement with the results of
\citet{holman_etal06}.

\section{Conclusions}
Here, we present new ground based high cadence near-infrared
observations of one transit of the hot-Neptune GJ\,436b, and four
transits, spanning three epochs, of the hot-Jupiter XO-1b.

We achieve transiting timing accuracies of about 30\,sec for
individual transits. The uncertainty is dominated by systematic
effects, and greatly exceeds the few second errors predicted by photon
noise dominated observations. We find no significant evidence for
perturbations of the orbital motion of GJ\,436b nor XO-1b by other
bodies in the system. Of course, a proper test of this hypothesis will
require monitoring of multiple transits with the same or even higher
accuracy.

We demonstrate that the ground-based high-cadence observations of
transiting extrasolar planets is an excellent technique for
constraining the parameters of extrasolar planetary systems because of
the statistical significance of the obtained timing measurements. The
timing precision is comparable with the space-based observations,
making this method a good alternative to the space mission with their
high cost and limited life-time.

\begin{acknowledgements}
We greatly acknowledge the ESO Director's Discretionary Time Committee
for the prompt response to our observing time request. DM and CC are
supported by the Basal Center for Astrophysics and Associated
Technologies, and the FONDAP center for Astrophysics 15010003.
\end{acknowledgements}

\bibliographystyle{aa}
\bibliography{paperI}

\end{document}